\documentclass[twocolumn,prl,showpacs,superscriptaddress,floatfix]{revtex4}
\usepackage{color}
\usepackage{colordvi}
\usepackage{amsmath}
\usepackage{amssymb}
\usepackage{fancybox}
\usepackage{epsfig}
\usepackage{graphicx}
\begin{document}

\title{Interacting lattice electrons with disorder  in two dimensions: Numerical evidence for a metal-insulator transition with a universal critical conductivity}

\author{Prabuddha B. Chakraborty}
\email{prabuddha.chakraborty@physik.uni-augsburg.de}
\affiliation{ Theoretical Physics III, Center for Electronic
Correlations and Magnetism, Institute for Physics, University of
Augsburg, D-86135, Augsburg, Germany}
\author{Krzysztof Byczuk}
\affiliation{ Faculty of Physics, University of Warsaw,
ul.  Ho\.za 69, 00-681, Warszawa, Poland}
\author{Dieter Vollhardt}
\affiliation{ Theoretical Physics III, Center for Electronic
Correlations and Magnetism, Institute for Physics, University of
Augsburg, D-86135, Augsburg, Germany}

\begin{abstract}

The dc-conductivity of electrons on a square lattice interacting with a local repulsion in the presence of disorder is computed  by means of quantum Monte Carlo simulations. We provide evidence for the existence of a  transition from an Anderson insulator to a correlated disordered metal with a universal value of the critical dc-conductivity $\sigma_{{\rm{dc,crit}}}=\left(1.18\pm 0.06\right)e^{2}/h$ at the transition.

\end{abstract}
\date{\today}

\maketitle

The Coulomb interaction between the electrons and the presence of disorder both strongly affect the properties of solids \cite{Lee85,Altshuler85,Belitz94,Abrahams01,Kravchenko04}. Namely, electronic correlations and
randomness are separately driving forces behind metal--insulator transitions
(MITs) due to the localization and delocalization of
particles. While the electronic repulsion may lead to a Mott-Hubbard
MIT \cite{Mott90}, the scattering  of
non-interacting particles from randomly distributed impurities can
cause Anderson localization  \cite{Anderson58,Anderson50-2010}.
The simultaneous presence of disorder and interactions lead to further subtle many-body effects
which raise fundamental questions in theory and experiment not only
in solid state physics
\cite{Finkelshtein83,Castellani84,Lee85,Altshuler85,Mott90,Logan93,Belitz94,Shepelyansky94,Abrahams01,Kravchenko04,Anderson50-2010}, but also in the field of cold atoms in optical lattices \cite{Cold_atoms}.

According to the scaling theory of Anderson localization \cite{Abrahams79,Gorkov79} non-interacting electrons in two spatial dimensions ($d=2$) are localized in the
presence of disorder. Hence, in the thermodynamic limit at zero temperature, there is
no metallic state in $d=2$.
By contrast, the experimental observation of a metal-insulator transition (MIT) in resistivity measurements on various high mobility heterostructure samples and Si-MOSFETs \cite{Kravchenko:1994} clearly indicates that interactions can turn an Anderson insulator into a metal. Near-perfect scaling of the resistivity data \cite{Kravchenko:1995} was taken as evidence for the presence of a quantum critical point between the metallic and the Anderson localized state \cite{Abrahams01,Kravchenko04}. Recent investigations \cite{Punnoose:2005,Anissimova:2007} based on a non-linear sigma model (NL$\sigma$M) for interacting electrons with disorder in the continuum confirm the existence of a such quantum critical point
which is characterized by a universal value of the dc-resistivity.
Universal critical conductivities were also discussed in other two-dimensional systems, e.g., in connection with  the transition from a superconductor (superfluid) to an insulator \cite{Fisher90,Goldman98}, in the integer quantum Hall effect \cite{Schweitzer05}, and in graphene 
 \cite{Ostrovsky07}.

Numerical investigations of the interplay between disorder and interactions usually address electrons on a lattice rather than in the continuum. Various approaches include Hartree-Fock investigations in $d=3$ \cite{Tusch:1993} and $d=2$ \cite{Heidarian:2004}, quantum Monte Carlo (QMC) simulations \cite{Denteneer:1999,Denteneer:2003,Chakraborty:2007}, and dynamical mean-field theory \cite{Dobrosavljevic:1993,Dobrosavljevic:2003,Byczuk:2005,Semmler:2010}. In their QMC studies of two-dimensional lattice electrons Denteneer {\emph{et al.}} \cite{Denteneer:1999, Denteneer:2003} indeed found a phase transition between an Anderson insulator and a metallic phase in accordance with experiment \cite{Kravchenko04}. There has also been the proposal of the MIT as a percolation transition \cite{Tracy:2009}.

In this Letter we provide evidence through extensive QMC
simulations that in the Anderson-Hubbard model in $d=2$  there exists a transition between a
metallic phase and an Anderson insulator, and that  this transition takes place at a value of the dc-conductivity $\sigma_{{\rm{dc,crit}}}$ which is essentially independent of the   critical interaction,  critical disorder,  and particle density. The computation of such a universal value of the critical dc-conductivity provides an explicit link to results obtained from effective theories in the continuum \cite{Punnoose:2005}. Indeed, numerical investigations of microscopic lattice models can provide details of the properties of a system at a quantum critical point which are not accessible within effective perturbative approaches.

Our investigation of interacting electrons in the presence of disorder is based on the Anderson-Hubbard Hamiltonian on a square lattice
\begin{equation}
H = T\{\epsilon_i\}+U\sum_i n_{i\uparrow}n_{i\downarrow}.
\label{Hamiltonian}
\end{equation}
Here
\begin{eqnarray}
T\{\epsilon_i\}&=&-t\sum_{<ij>\sigma}
c_{i\sigma}^\dagger c_{j\sigma}
+\sum_{i \sigma}(\epsilon_i - \mu)n_{i\sigma},
\label{quadratic}
\end{eqnarray}
is the single-electron part where $c_{i\sigma}^\dagger$, ($c_{i\sigma}$) are
fermion creation (annihilation) operators for site $\textbf{R}_i$ and spin $\sigma$,
$n_{i\sigma}=c_{i\sigma}^\dagger c_{i\sigma}$ is the operator for the local density, $\mu$ denotes the chemical potential, and $t$ is the hopping amplitude for electrons between nearest neighbor sites.
%
The local energies $\epsilon_i$ are random variables which are sampled uniformly from the
interval $[-\Delta/2,\Delta/2]$; hence the width $\Delta$ characterizes the strength of the disorder. The interaction is assumed to be repulsive ($U>0$).
The model is solved numerically using determinantal QMC (DQMC) \cite{Blankenbecler:1981} where the interval $[0,\beta]$ $\left(\beta=1/k_BT\right)$ is partitioned according to $\beta=L\times\Delta\tau$, with $\Delta\tau$ as the size of a small step in the imaginary time direction, and $L$ as the number of imaginary time slices. The partition function $Z$ is then decomposed according to the Suzuki-Trotter formula \cite{Suzuki:1976}.
In the next step, a Hubbard-Stratonovich transformation is performed whereby the interaction problem is reduced to  non-interacting electrons in the presence of infinitely many fluctuating fields described by Ising variables on every space-(imaginary) time lattice site \cite{Hirsch:1985}. The electrons can then be integrated out. The calculation of quantities such as the Green function, electronic density and two-particle correlation functions proceeds with Monte Carlo sampling of the various configurations of the Ising degrees of freedom. The hopping integral $t$ sets the unit of energy and the simulation now contains three independent energy-scales:
the disorder strength $\Delta$, the interaction strength $U$, and the temperature $T$.

To evaluate the dc-conductivity, we compute the electronic current density operator
\begin{equation}
j_{x}(\textbf{R}_{i})=\frac{ieat}{\hbar}\sum_{\sigma}\left( c_{i+\textbf{e}_x \sigma}^{\dagger}c_{i \sigma}-c_{i \sigma}^{\dagger}c_{i+ \textbf{e}_x \sigma} \right),
\end{equation}
where $\textbf{e}_x$ denotes a translation in $x$-direction by a lattice constant $a$. This leads to the time-dependent current density operator
\begin{equation}
j_{x}(\textbf{R}_{i},\tau)=e^{H\tau/\hbar}j_{x}(\textbf{R}_{i})e^{-H\tau/\hbar},
\end{equation}
where $\tau$ is the imaginary (Matsubara) time.
The position-space Fourier transform of the current operator, $j_{x}(\textbf{q},\tau)$, is then used to calculate the current-current correlation function
\begin{equation}
\Lambda_{xx}(\textbf{q},\tau)=\langle j_{x}(\textbf{q},\tau)j_{x}(-\textbf{q},\tau=0) \rangle.
\end{equation}
Within linear response theory, the dc-conductivity is obtained from
\begin{equation}
\sigma_{\rm{dc}}=\lim\limits_{\omega\rightarrow 0}\frac{{\rm{Im}}\Lambda_{xx}(\textbf{q}=0,\omega)}{\omega}.
\label{linres}
\end{equation}
The current-current correlation function in Matsubara time is related to the imaginary part of the current-current correlation function
in real frequency through the integral transform
\begin{equation}
\Lambda_{xx}(\textbf{q},\tau)=\int_{-\infty}^{\infty}\frac{d\omega}{\pi}\frac{e^{-\omega\tau}}{1-e^{-\beta\omega}}{\rm{Im}}\Lambda_{xx}(\textbf{q},\omega).
\label{kernel}
\end{equation}
DQMC simulations can compute $\Lambda_{xx}(\textbf{q},\tau)$, but to determine $\sigma_{\rm{dc}}$ it is necessary to obtain ${\rm{Im}}\Lambda_{xx}(\textbf{q},\omega)$. For low enough temperatures, the exponential decay of the bosonic kernel $K(\omega,\tau;\beta)=\frac{e^{-\omega\tau}}{1-e^{-\beta\omega}}$ for $\tau = \beta/2$ ensures that the integral contributes only for small $\omega$, where the substitution arising from linear response, eq.~(\ref{linres}), is valid. Replacing $\tau$ by $\beta/2$ and ${\rm{Im}}\Lambda_{xx}(\textbf{q},\omega)$ by $\omega\sigma_{{\rm{dc}}}$, the integral can be carried out analytically and yields the dc-conductivity \cite{Randeria:1992} as a function of temperature for different values of the interaction $U$ and disorder strength $\Delta$:
\begin{equation}
\sigma_{{\rm{dc}}}=\frac{\beta^2}{\pi}\Lambda_{xx}\left(\textbf{q}=0,\tau=\frac{\beta}{2}\right).
\label{finalformula}
\end{equation}
In the following discussion, we set $t=1$.

The conductivity data is averaged over 10 disorder realizations at high temperatures ($T=1,0.5,0.333,0.25$), up to 80 disorder realizations for intermediate temperatures ($T=0.2,0.167$), and
up to 100 disorder realizations for the two lowest temperatures ($T=0.125,0.1$). In
\begin{figure}
\includegraphics[clip=true,angle=-0,width=0.45\textwidth]{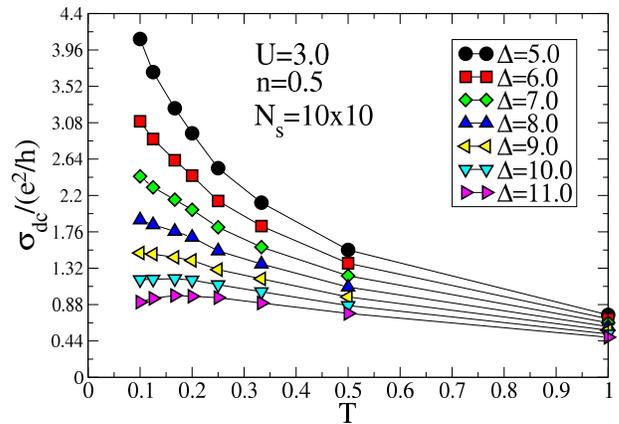}
\caption{(Color online) Curves of the dc-conductivity \emph{vs.} temperature  at electron density $n=0.5$ (quarter filling)
and interaction $U=3$ on a $10\times10$ square lattice computed for different values of the disorder strength $\Delta$ (see inset).}
\label{sigman0_5U3_scanD}
\end{figure}
Fig.~\ref{sigman0_5U3_scanD}, the dc-conductivity is shown as a function of T for several values of the disorder strength $\Delta$. Initially, when the value of the
disorder strength is less than about $\Delta=10$, the slope of the
conductivity curve at low temperatures is negative (i.e., the conductivity decreases with increasing temperature), implying that the system is
metallic. As the disorder strength is increased, the low temperature conductivity develops a positive slope, which is the
signature of an insulator. Since the system is far from half-filling, such that a Mott-Hubbard MIT does not occur, these results indicate a transition between a metallic and
an Anderson localized state.

\begin{figure}
\includegraphics[clip=true,angle=-0,width=0.45\textwidth]{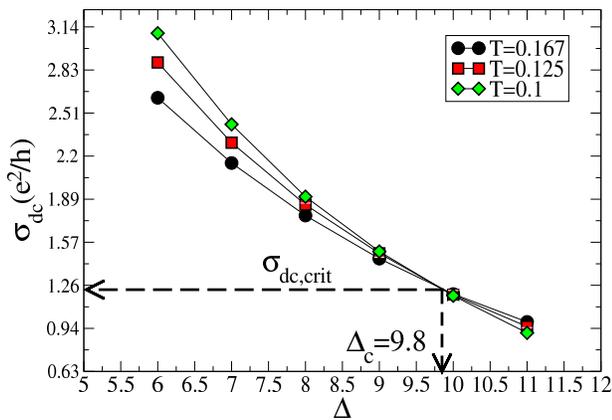}
\caption{(Color online) Plot of the critical conductivity
$\sigma_{{\rm{dc}}}$ \emph{vs.}
disorder strength $\Delta$ for three temperatures and the same values of $U$ and $n$ as in Fig.~\ref{sigman0_5U3_scanD}. The well-defined crossing point determines the critical disorder strength as $\Delta_{{\rm{c}}}=9.8$ . The  conductivity at the critical disorder has the value $\sigma_{{\rm{dc,crit}}}=1.19 e^2/h$.}
\label{crossingn0_5U3}
\end{figure}

There are two sources of statistical error in this analysis: one due to  the QMC simulations, the other due to the disorder averaging. For all parameter sets
studied here, the intrinsic QMC error for any given disorder realization is much smaller than the error arising from different disorder realizations. Since the error bars in Figs.~\ref{sigman0_5U3_scanD}, \ref{crossingn0_5U3}
are of the order of, or smaller, than the symbols they are not shown.

\begin{figure}
\includegraphics[clip=true,angle=-90,width=0.45\textwidth]{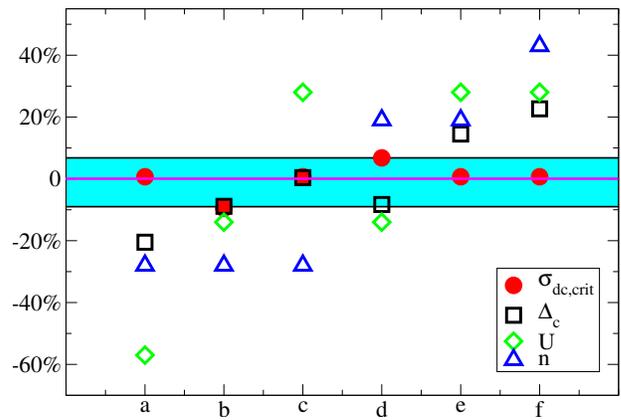}
\caption{(Color online) Graphical presentation of the spread of the parameters listed in Table~\ref{sets} in percentage relative to $\tilde{p}$, where $p$ is one of the quantities $U$, $n$, $\Delta_{{\rm{c}}}$ and $\sigma_{{\rm{dc,crit}}}$. Here $\tilde{p}$ defines the arithmetic mean of the parameter $p$ across the six parameter sets a-f. For example, $\tilde{U}= \frac{1}{6}\left(1.0+2.0+3.0+2.0+3.0+3.0\right)=2.33$. The quantity $\tilde{U}$ does not have a physical meaning, but is only a reference point to display the relative spread of the $U$ values used in the six parameter sets; the same holds for the parameters $\tilde{n}$ and $\tilde{\Delta_{{\rm{c}}}}$. By contrast,  $\sigma_{{\rm{dc,crit}}}$ varies very little across the parameter sets and clusters around $1.18e^2/h$. Therefore the quantity $\tilde{\sigma}_{{\rm{dc,crit}}}$ can be accorded physical meaning. }
\label{spread}
\end{figure}

On the basis of Fig.~\ref{sigman0_5U3_scanD} neither the critical disorder strength nor the value of the dc-conductivity $\sigma_{{\rm{dc}}}$ at the critical point can be determined accurately. In Fig.~\ref{crossingn0_5U3} we therefore plot $\sigma_{{\rm{dc}}}$ as a function of the disorder $\Delta$ for the three lowest temperatures simulated here, i.e., $T=0.167,0.125,0.1$. When $\Delta < \Delta_{{\rm{c}}}$, the dc-conductivity increases with decreasing temperature (metallic behavior), while for $\Delta > \Delta_{{\rm{c}}}$, the conductivity decreases with decreasing temperature (insulating behavior). The three curves shown in Fig.~\ref{crossingn0_5U3}
display a well-defined crossing point, at which the dc-conductivity is independent of temperature, thereby marking the critical point for the MIT. From the location of the crossing point one can read off the value of the critical disorder strength $\Delta_{{\rm{c}}}$. For $U=3.0$ at quarter-filling ($n=0.5$) we find $\Delta_{{\rm{c}}}=9.8$, while the value of the critical conductivity is $\sigma_{{\rm{dc,crit}}}=1.19e^2/h$. We will use this technique to evaluate the critical disorder strength and the critical conductivity for five more parameter sets $(U,n)$ listed in Table~\ref{sets}.
\begin{table}[tbp!]
\caption{The six parameter sets of the interaction strength $U$ and electron density $n$ employed in our investigation of correlated electrons in the presence of disorder on a square lattice are listed together with the computed critical disorder strength $\Delta_{{\rm{c}}}$ and the critical dc-conductivity $\sigma_{{\rm{dc,crit}}}$ at the transition between a disordered metal and an Anderson insulator.}
\begin{ruledtabular}
\begin{tabular}{lcccc}
Index & $U$ & $n$  & $\Delta_{{\rm{c}}}$ & $\sigma_{{\rm{dc,crit}}}$ ($e^2/h$) \\
\hline
a                    & 1.0  & 0.3   &  6.8  &  1.19 \\
b                    & 2.0  & 0.3   &  7.8  &  1.07 \\
c                    & 3.0  & 0.3   &  8.6  &  1.19 \\
d                    & 2.0  & 0.5   &  7.9  &  1.26 \\
e                    & 3.0  & 0.5   &  9.8  &  1.19 \\
f                    & 3.0  & 0.6   & 10.5  &  1.19 \\
\end{tabular}
\end{ruledtabular}
\label{sets}
\end{table}
The results collected in this table can be summarized as follows. In spite of the strong variation of the microscopic input parameters $U,n$ and the disorder strength $\Delta_{{\rm{c}}}$ at the transition between the metallic and the Anderson-insulating state in $d=2$, the associated critical dc-conductivity, $\sigma_{{\rm{dc,crit}}}$, is found to be essentially independent of these input parameters. The results are presented graphically in Fig.~\ref{spread}, where the value of the critical conductivity is seen to cluster around the value $\sigma_{{\rm{dc,crit}}}=\left(1.18 \pm 0.06\right)e^2/h$.
%
This provides evidence for the existence of a \emph{universal} value of the critical conductivity. Indeed, the  NL$\sigma$M, in which the dc-conductivity appears as a coupling constant, predicts a universal \cite{Spivak:2010} critical value of the conductivity $\sigma_{{\rm{dc,crit}}}\sim 1.06e^2/h$ \cite{Punnoose:2005}, in close correspondence with our result. In obtaining this estimate, we assumed that the number of valleys appropriate for our work is $n_v=1$  \cite{v=1}.
Thus our results establish a link between the microscopic Anderson-Hubbard model, and the low-energy effective theory provided by the NL$\sigma$M for the metal-insulator phase transition in two dimensions.

In summary, quantum Monte-Carlo simulations of interacting lattice electrons in the presence of disorder in $d=2$ provide clear evidence for a transition from metallic to insulating behavior as the disorder strength is varied. At the transition the value of the dc-conductivity is found to be given by $\sigma_{{\rm{dc,crit}}}=\left(1.18 \pm 0.06\right)e^2/h$, implying that the critical dc-conductivity is essentially independent of interaction strength, electron density and the critical disorder strength. This points towards the existence of a universal critical dc-conductivity. We obtained qualitatively similar results from investigations where site-disorder is changed to bond-disorder, the details of which will be published elsewhere.

This work was supported in part by the SFB 484 (2000-2009) of the Deutsche Forschungsgemeinschaft and the TRR 80. One of us (KB) also acknowledges support through the grant N~N202~103138 of the Polish Ministry of Science and Education.  We are grateful for valuable discussions and communications with N. Trivedi, R. T. Scalettar, S. Chiesa, A. M. Finkel'stein, and J. Tworzyd{\l}o.

\end{document}